\documentclass[english,prl,twocolumn,amsmath,amssymb,superscriptaddress,floatfix]{revtex4}

\usepackage[T1]{fontenc}
\usepackage[latin1]{inputenc}
\usepackage{graphicx}
\usepackage{float}
\usepackage{epstopdf}
\usepackage{babel}
\makeatother

\begin{document}

\title{Multiplexed Charge-locking Device for Large Arrays of Quantum Devices}

\author{R.K. Puddy}
\altaffiliation{rkp27@cam.ac.uk}
\affiliation{Cavendish Laboratory, University of Cambridge, Cambridge CB3 0HE, United Kingdom}
\author{L.W Smith}
\affiliation{Cavendish Laboratory, University of Cambridge, Cambridge CB3 0HE, United Kingdom}
\author{H. Al-Taie}
\affiliation{Cavendish Laboratory, University of Cambridge, Cambridge CB3 0HE, United Kingdom}
\affiliation{Centre for Advanced Photonics and Electronics, Electrical Engineering Division, Department of Engineering, 9 J. J. Thomson Avenue, University of Cambridge, Cambridge, CB3 0FA, United Kingdom}
\author{C. H. Chong}
\affiliation{Cavendish Laboratory, University of Cambridge, Cambridge CB3 0HE, United Kingdom}
\author{I. Farrer}
\affiliation{Cavendish Laboratory, University of Cambridge, Cambridge CB3 0HE, United Kingdom}
\author{J.P. Griffiths}
\affiliation{Cavendish Laboratory, University of Cambridge, Cambridge CB3 0HE, United Kingdom}
\author{D.A.  Ritchie}
\affiliation{Cavendish Laboratory, University of Cambridge, Cambridge CB3 0HE, United Kingdom}
\author{M.J. Kelly}
\affiliation{Cavendish Laboratory, University of Cambridge, Cambridge CB3 0HE, United Kingdom}
\affiliation{Centre for Advanced Photonics and Electronics, Electrical Engineering Division, Department of Engineering, 9 J. J. Thomson Avenue, University of Cambridge, Cambridge, CB3 0FA, United Kingdom}
\author{M. Pepper}
\affiliation{Deptartment of Electronic and Electrical Engineering, University College London, UK.}
\author{C.G. Smith}
\affiliation{Cavendish Laboratory, University of Cambridge, Cambridge CB3 0HE, United Kingdom}

\begin{abstract}

We present a method of forming and controlling large arrays of gate-defined quantum devices. The method uses a novel, on-chip, multiplexed charge-locking system and helps to overcome the restraints imposed by the number of wires available in cryostat measurement systems. Two device innovations are introduced. Firstly, a multiplexer design which utilises split gates to allow the multiplexer to divide three or more ways at each branch. Secondly we describe a device architecture that utilises a multiplexer-type scheme to lock charge onto gate electrodes. The design allows access to and control of gates whose total number exceeds that of the available electrical contacts and enables the formation, modulation and measurement of large arrays of quantum devices. We fabricate devices utilising these innovations on n-type GaAs/AlGaAs substrates and investigate the stability of the charge locked on to the gates. Proof-of-concept is shown by measurement of the Coulomb blockade peaks of a single quantum dot formed by a floating gate in the device. The floating gate is seen to drift by approximately one Coulomb oscillation per hour.

\end{abstract}

\maketitle

Motivation for the measurement of large numbers of quantum devices arises both from interest in the associated physical properties such as the formation of minibands\cite{miniband} and from the drive to up-scale and integrate quantum phenomenon, such as spin physics\cite{hanson spins}, into future technology and quantum information processing\cite{loss}. Much of the physics of interest is only observable using cryogenic systems and the number of coupled devices is limited by the number of available contacts. Recent work has shown the use of multiplexing to greatly increase the number of isolated quantum devices available for measurement on a single chip and single cool-down\cite{haider1,mux dbl dots} and frequency multiplexing, for the readout of spin qubits\cite{freqMUX}, has been demonstrated as a potential up-scaling route. The significant challenges presented by the need to up-scale however are far from surmounted.\\
\indent The measurement of many individually addressable quantum devices has led to initial studies on yield\cite{haider1}, reproducibility\cite{haider2repro} and statistical analysis of complex quantum phenomena\cite{LukePRB}. The split gate\cite{sgpaperfirst,sgpaperquant1,sgpaperquant2} can be considered as the building block for more complex gate-defined devices, such as quantum dots\cite{cgs}. Tuneable quantum dots require stable charge on several surface gates simultaneously in order to function. The multiplexing architecture presented in \cite{haider1} doesn't allow the simultaneous use of multiple gates. We therefore present two innovations that facilitate the fabrication and measurement of large interacting quantum device arrays.
\begin{figure}
\includegraphics[width=85mm]{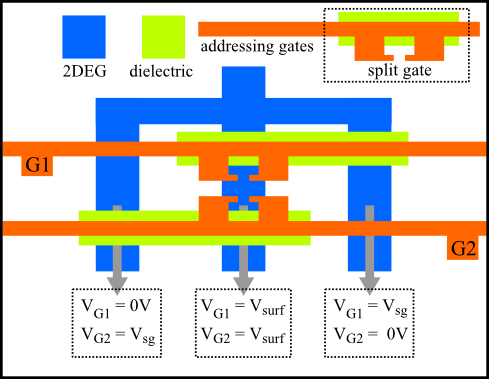}
\caption{\label{fig3way} Schematic of a novel multiplexer geometry using a split gate to enable three-fold branching. Addressing gates G1 \& G2 (red) pass over the left and right 2DEG branches (blue) either close to the 2DEG, or separated by a dielectric (green). Both G1 \& G2 form  split gates above the central channel. Split gate dimensions are chosen such that the surface gate pinch off voltage V$_{\textrm{surf}}$ has only a small effect on the 2DEG conductance under the split gate and the thickness of the  dielectric is such that the 2DEG under the dielectric is affected only negligibly by V$_{\textrm{sg}}$ or V$_{\textrm{surf}}$. The voltages applied to gates G1 \& G2, required for addressing each of the three channels are given in the dashed boxes below the relevant outputs.}
\end{figure}
\indent We firstly show how a split gate can be used within a multiplexer-type addressing system, to enable the multiplexer to divide three or more ways at each node rather than two. Figure \ref{fig3way} shows a schematic of a single node of a 3-way multiplexer. A semiconducting two dimensional electron gas (2DEG), shown in blue, divides into three channels. Addressing-gates, G1 and G2, shown in red, pass over the left and right channels, either directly over the substrate surface, or separated by a dielectric, shown in green. Both G1 and G2 form split gates above the central channel. We define two voltages, the split gate pinch off V$_{\textrm{sg}}$ and the surface gate pinch off V$_{\textrm{surf}}$. The width of the split gate and the thickness of the dielectric are chosen such that applying V$_{\textrm{surf}}$ to the addressing gates has only a small effect on the conductance of the 2DEG under the split gate and applying either V$_{\textrm{surf}}$ or V$_{\textrm{sg}}$ has a negligible effect on the conductance of the 2DEG below the dielectric. The voltage combinations required to address each channel are given in the dashed boxes below the relevant output. Applying V$_{\textrm{sg}}$ to G2 closes the right and central channels. Applying V$_{\textrm{surf}}$ simultaneously to G1 and G2 closes the left and right channels. Applying V$_{\textrm{sg}}$ to G1 closes the left and central channels. Such a multiplexer therefore provides 3$^{(n-1)/2}$ outputs from a single input using $n$ contacts.
\begin{figure}[H]
\includegraphics[width=85mm]{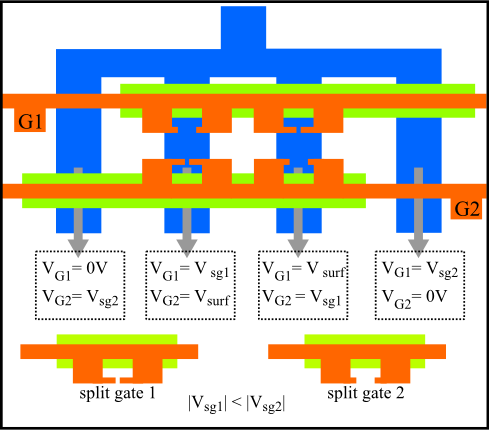}
\caption{\label{fig4way} Schematic of a multiplexer that uses split gates of two different widths in order to allow four-way branching. The voltage combinations applied to G1 and G2 in order to address a given output are shown in the dashed boxes. Two different widths of split gate are used giving two distinct pinch off voltages, V$_{\textrm{sg1}}$ and V$_{\textrm{sg2}}$ with $|V_{\textrm{sg1}}|<|V_{\textrm{sg2}}|$.}
\end{figure}
The principle described above can be extended to construct multiplexers with four or more branches by incorporating further split gates. In figure \ref{fig4way} we show a multiplexer with 4-fold branching. Here, two split gate widths are used, giving two distinct pinch off voltages $|V_{sg1}|<|V_{sg2}|$. The gate voltage combinations, applied to the addressing gates G1 and G2, required to address any given output are shown in the dashed boxes below the relevant output. More branching is possible as long as split gates can be fabricated with pinch off voltages that are sufficiently different from each other.

\begin{figure}
\includegraphics[width=85mm]{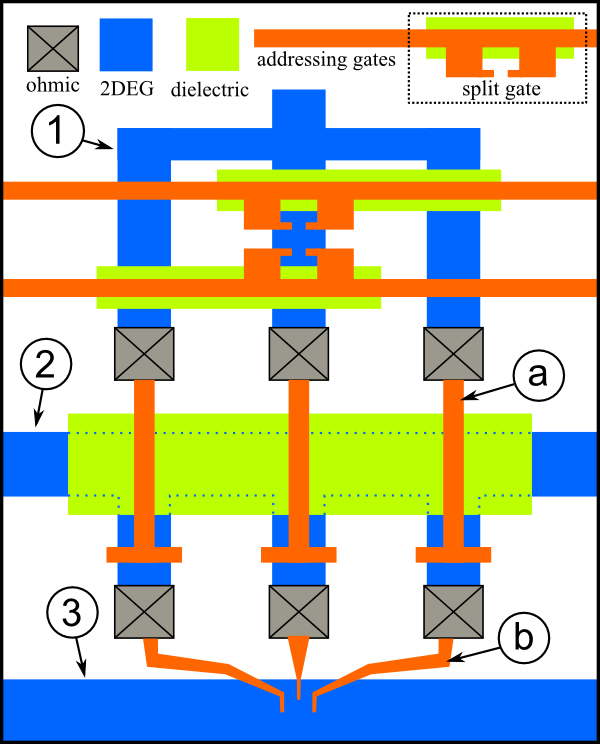}
\caption{\label{fig3waydevice} Diagram of a charge-locking device consisting of three separate 2DEGs (blue) which we denote as: \textbf{(1)} the multiplexer 2DEG, \textbf{(2)} the gate-source 2DEG and \textbf{(3}) the measurement 2DEG. The multiplexer, operated by addressing gates, has multiple ohmic outputs (grey with black cross) terminating in surface electrodes, the locks \textbf{(b)}. The locks cross a layer of dielectric and cover the gate-source tributaries. A dielectric protects the main channel of the gate-source 2DEG. The gate-source 2DEG, under the dielectric is highlighted by the blue dashed line. Ohmic contacts at the outputs of the gate-source 2DEG connect to surface gates that form the device to be measured on the measurement 2DEG.}
\end{figure}

\begin{figure}
\includegraphics[width=85mm]{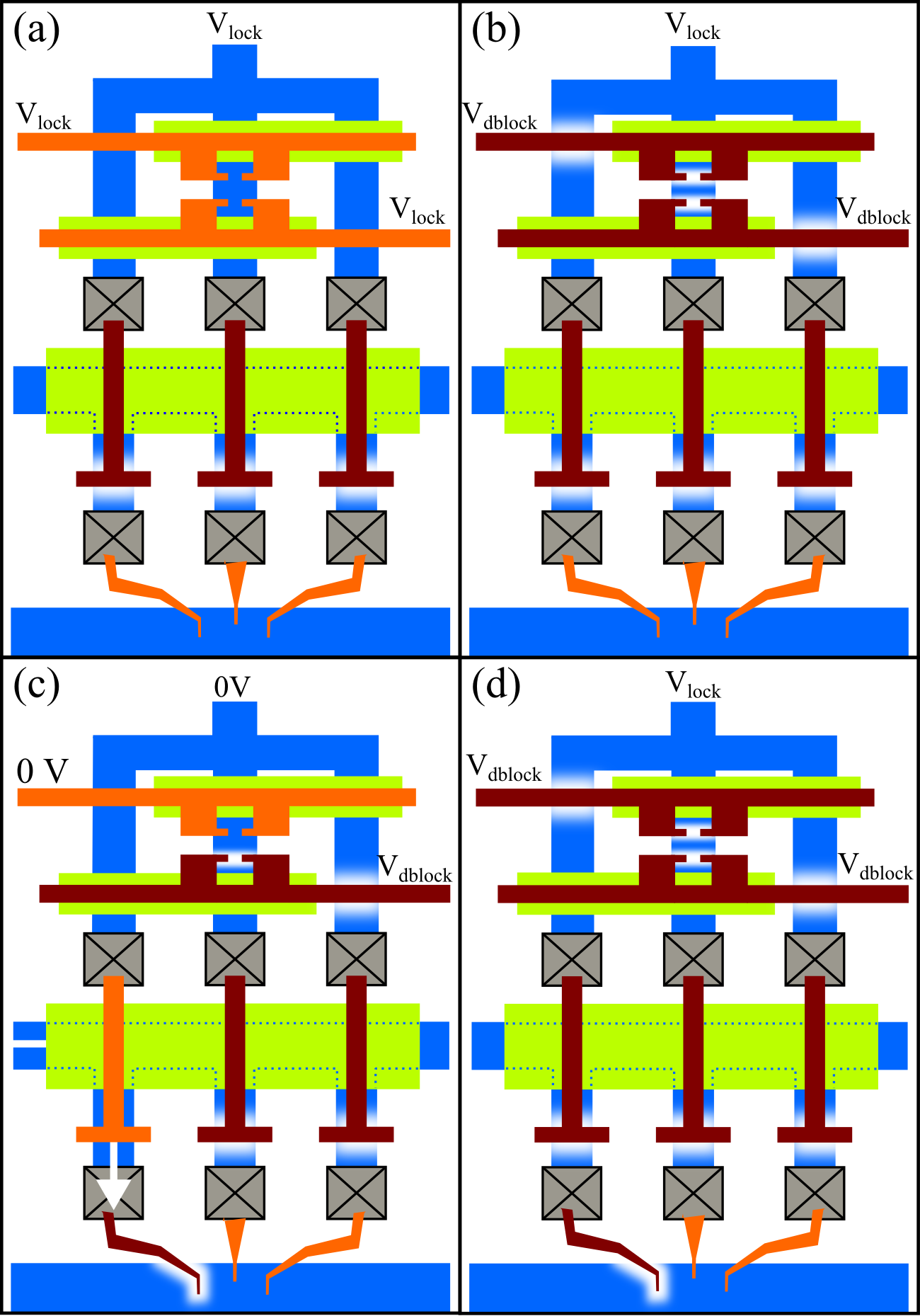}
\caption{\label{fig3operation} Schematic showing the four basic stages of operation of our charge-locking device. \textbf{(a)} the multiplexer 2DEG and addressing gates are set to a voltage well beyond the depletion voltage, e.g. -1 V, closing all the tributaries of the gate-source 2DEG. \textbf{(b)} The addressing gates are set to a voltage above the split gate pinch off voltage, say -3 V, leaving the locking-gates charged and isolated. \textbf{(c)} The left locking-gate is addressed and the multiplexer set to 0 V. This releases charge trapped on lock 1 so that the left most device-gate is connected to the input of the device-gate 2DEG and the voltage can be swept to the desired value. \textbf{(d)} The multiplexer 2DEG and locking-gate is set to -1 V, locking the charge onto device-gate 1 and then the addressing gate is set to -3 V isolating the lock. The procedure is then repeated for the next device-gate.}
\end{figure}
 
We next describe a device geometry that utilises a multiplexer to lock charge onto surface-gate electrodes. A diagram of the device is shown in figure \ref{fig3waydevice}. The example shown here uses our 3-way multiplexer. The device consists of three separate 2DEGs which we denote as (1) the multiplexer 2DEG, (2) the gate-source 2DEG and (3) the measurement 2DEG. The multiplexer outputs connect to surface gate electrodes, (a) in the figure, referred to as locking-gates. The gate-source 2DEG consists of a comb-like structure with a main channel and multiple tributaries. The main channel is covered by a dielectric, the structure of the gate-source 2DEG under the dielectric is picked out by the blue dotted lines. Each tributary is connected to a surface electrode referred to as a device-gate, these pass onto the surface of the measurement 2DEG to define the system to be measured. Each locking-gate passes over the dielectric and covers a single tributary of the gate-source 2DEG. The stages of operation of the device are shown in figure \ref{fig3operation}. Firstly, figure \ref{fig3operation} (a), the multiplexer 2DEG and addressing gates are set to a voltage which we name the locking voltage $V_{\textrm{lock}}$, that is well beyond the depletion voltage of the 2DEG V$_{\textrm{surf}}$ (active gates are coloured maroon). This initial operation depletes the 2DEG under the locking-gates (depletion is represented by white blurring in the figure) thus isolating all the gate-source tributaries from the main channel. Next, figure \ref{fig3operation} (b), the addressing gates are set to a voltage, which we name the double lock voltage $V_{\textrm{dbllock}}$, which is well beyond the split gate pinch-off voltage, V$_{\textrm{sg}}$. This second operation isolates the locking-gates which are now charged and floating at $V_{\textrm{lock}}$. We next address one of the multiplexer outputs, e.g. the left branch in figure \ref{fig3operation} (c), and set the multiplexer 2DEG to 0V. The addressed locking-gate discharges and the tributary is thus re-connected to the main channel of the gate-source 2DEG. This allows a single device-gate to be swept to the desired voltage. Next, figure \ref{fig3operation} (d), the locking-gate is set to $V_{\textrm{lock}}$ via the multiplexer input to isolate and lock the charge onto the device-gate, and the addressing gates are set to $V_{\textrm{dbllock}}$ to isolate the locking-gates. The operations in figure \ref{fig3operation} (b)-(c) can then be repeated for the other device-gates. In this way large numbers of gates can be set up to form complex devices.

\begin{figure}
\includegraphics[width=85mm]{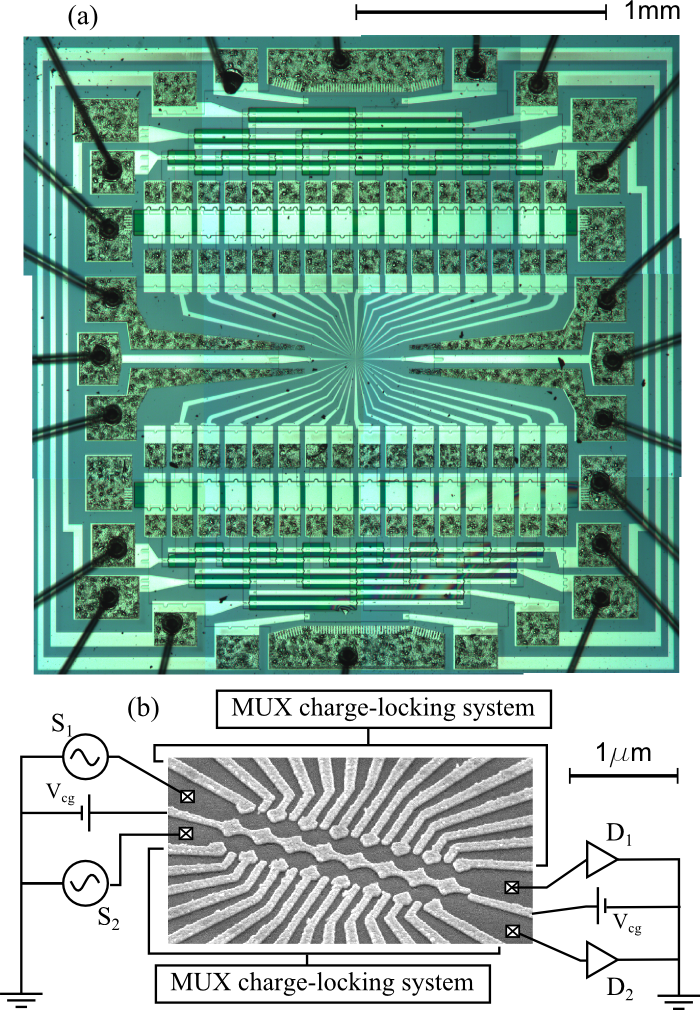}
\caption{\label{optical image device} \textbf{(a)} Optical image of charge-locking device, fabricated on GaAs/AlGaAs, consisting of two opposing multiplexed charge-locking systems. The multiplexers are of 2-way design with four addressing levels (16 outputs). The first three addressing gate levels are shared to save contact numbers. \textbf{(b)} SEM image of the central region of the measurement 2DEG and wiring. The measurement 2DEG is divided in two by a central gate electrode directly controlled by a voltage supply, $V_{\textrm{cg}}$, such that two measurement channels are available with two source and two drain contacts, S$_{1,2}$ and D$_{1,2}$ respectively. The central gate is made up of two separately contactable parts with a 20nm gap to add an extra level of control. Fifteen gates are used from each of the multiplexer units and are arranged so that they can form two parallel rows of seven QDs.}
\end{figure}

\begin{figure}
\includegraphics[width=85mm]{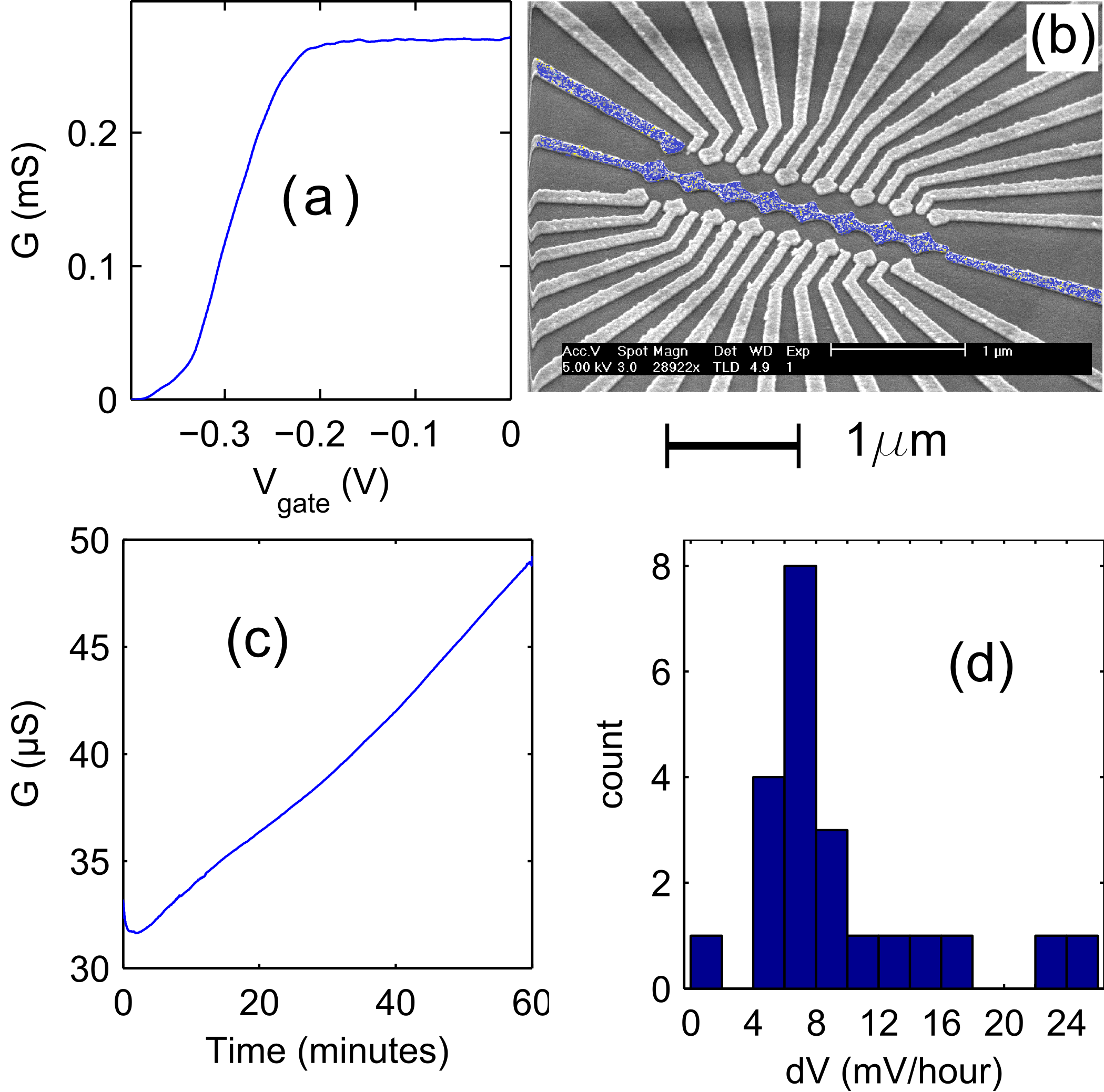}
\caption{\label{figholding} Measurements of the charge-locking device shown in  Fig.\ref{optical image device} carried out at $\approx$ 50 mk. \textbf{(a)} Conductance as a function of device gate voltage of a narrow channel defined by the central gate and a single device gate (highlighted in blue in the SEM image in panel \textbf{(b)}). The device gate is addressed as in Fig. \ref{fig3operation} (c), and directly connected to the gate source input. \textbf{(c)} Conductance as a function of time of a single floating device-gate. The device-gate has been isolated using the charge-locking multiplexer system, as in Fig. vjhvjh\ref{fig3operation} (d), and no external voltages are applied to the device-gate 2DEG. The time varying conductance, $dG/dt$, is converted to an effective change in device-gate voltage, $dV_{\textrm{g}}/dt$ by comparison of (c) with (a). Panel \textbf{(d)} shows a histogram of the calculated $dV_{g}/dt$ for several device-gates.}
\end{figure}

\begin{figure}
\includegraphics[width=85mm]{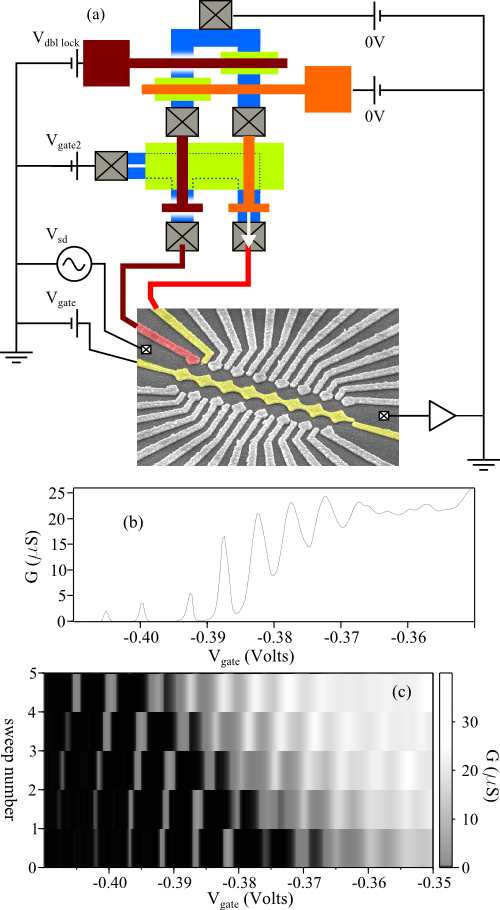}
\caption{\label{figqds} \textbf{(a)} Schematic of a proof-of-principle charge-locking device measurement. Gate 1 (highlighted dark red) has been addressed, set to -0.45 V and locked as in Fig. \ref{fig3operation} (d). Gate-2 is addressed as described in Fig. \ref{fig3operation} (c). The central gate is directly connected to a voltage supply and fixed at V$_{\textrm{gate}}$ = -0.5 V. \textbf{(b)} Conductance as a function of gate-2 voltage showing Coulomb blockade resonances. \textbf{(c)} Grey-scale plot of five gate-2 sweeps carried out at one hour intervals showing a drift of approximately one CB period per hour, $\approx$ 8mV/hr.}
\end{figure}

We next present measurements of a device fabricated on modulation doped GaAs/AlGaAs high electron mobility transistor substrate with a 2DEG 90nm below the surface. The device consists of two opposing charge-locking systems built around standard 2-way multiplexers each terminating in 16 outputs. We use a $\approx$ 600nm layer of polyimide as the dielectric. Optical and SEM images of the device are shown in figure \ref{optical image device} (a) and (b) respectively. Fifteen device gates are used from each side. These meet on the measurement 2DEG separated by a central gate. The central gate (split into two, separately contactable sections to give greater control) forms two measurement channels each with a source and drain contact. The device-gates are arranged so that they can form two parallel rows of seven QDs. The first three of the four addressing levels are shared to save contact numbers.  Minimum separation between the gate electrodes is $\approx$ 20 nm and the lithographically designed diameter of the dots is 300 nm. The device requires 20 contacts to enable full control. We first look at the stability of single floating gates. All the following measurements are carried out at $\approx$ 50 mK, the base temperature of our dilution refrigerator. The central gate is fixed at -0.5 V (2DEG depletion voltage $\approx$ -0.3 V) and a single gate is addressed as in figure \ref{fig3operation} (c). The conductance $G$ as a function of device-gate voltage $V_{\texttt{dg}}$ is then measured. Figure \ref{figholding} (a) shows a representative plot of G as a function of V$_{\texttt{dg}}$ for the gate highlighted in blue in (b). The gate is then isolated, as in figure \ref{fig3operation} (d), and the conductance monitored as a function of time. Figure \ref{figholding} (c) shows such a plot. By comparing the two plots, i.e. figure \ref{figholding} (a) and (c), the time varying conductance due to e.g. leakage from the charged, isolated device-gate can be converted in to an effective change in device-gate voltage. Figure \ref{figholding} (d) shows a histogram of the effective drift of several gates. The modal average is around 7 mV/hr. We next show proof-of-principle by forming a QD between the central gate, a floating device-gate and an addressed device-gate. A circuit diagram of such a measurement is shown in figure \ref{figqds} (a). The central gate is held at -0.5 V and device-gate 1 is set to -0.4 V and isolated. Device gate-2 is then addressed and the voltage swept. Figure \ref{figqds} (b) shows the Coulomb blockade resonances appearing as device-gate 2 is swept. This measurement is repeated five times with one hour intervals between each sweep. A grey-scale plot of this is shown in figure \ref{figqds} (c). The drift due to the floating gate is $\approx$ 8 mV, about one Coulomb period per hour for this dot.

The device and measurement circuit shown in figure \ref{optical image device} (b) allows in principle the formation of multiple QDS on one side of the central gate whilst single gates on the parallel channel may be activated to form a quantum point contacts for non-invasive sensing measurements\cite{qpc}. We envisage using such an arrangement for the study of phenomenon such as spin waves and miniband formation.

Investigation into the causes of gate instability is required, further improvements may be achievable with the use of top gates. However, the techniques and device designs presented here represent a potentially powerful tool for the up-scaling of quantum devices and for the investigation of the basic physics associated with large numbers of coupled devices.

This work was supported by the Engineering and Physical Sciences Research Council Grant No. EP/K004077/1. The authors would like to thank R. D. Hall for electron-beam exposure.

\end{document}